\documentclass[12pt,preprint]{aastex}

\slugcomment{}

\shorttitle{Emerging flux regions on the solar surface}
\shortauthors{Cheung}

\begin{document}


\title{Solar surface emerging flux regions: a comparative study of radiative MHD modeling and Hinode SOT observations}


\author{M. C. M. Cheung\affil{Lockheed Martin Solar and Astrophysics Laboratory, Palo Alto, CA 94304, USA.}}
\author{M. Sch\"ussler\affil{Max Planck Institute for Solar System Research, Katlenburg-Lindau, 37191, Germany.}}
\author{T. D. Tarbell and A. M. Title\affil{Lockheed Martin Solar and Astrophysics Laboratory, Palo Alto, CA 94304, USA.}}




\begin{abstract}
We present results from numerical modeling of emerging flux regions on the solar surface. The modeling was carried out
by means of 3D radiative MHD simulations of the rise of buoyant magnetic flux tubes through the convection zone and into the photosphere.
Due to the strong stratification of the convection zone, the rise results in a lateral expansion of the tube into a magnetic sheet, which acts as a reservoir for small-scale flux emergence events at the scale of granulation. The interaction of the convective downflows and the rising magnetic flux undulates it to form serpentine field lines emerging into the photosphere. Observational characteristics including the pattern of emerging flux regions, 
the cancellation of surface flux and associated high speed downflows, the convective collapse of photospheric flux tubes, the appearance of anomalous darkenings, the formation of bright points and the possible existence of transient kilogauss horizontal fields are discussed in the context of new observations
from the Hinode Solar Optical Telescope. Implications for the local helioseismology of emerging flux regions are also discussed.

\end{abstract}





\section{Introduction}

The most prominent magnetic structures on the solar surface are bipolar active regions. These magnetic complexes are comprised of a hierarchy of magnetic structures of different sizes and exist over a range of timescales. With unprecedented spatial resolution, time cadence and a stable point-spread-function, the Solar Optical Telescope~\citep[SOT,][]{Tsuneta:SOT} onboard Hinode~\citep[formerly known as Solar-B,][]{Kosugi:Hinode} is providing new insights into the flux emergence process. Using Stokes profiles taken with the SOT SpectroPolarimeter~\citep*[SP,][]{Lites:HinodeSP}, various studies have been carried out to examine magnetic flux emergence at scales of the surface granulation. Whereas~\citet{Centeno:SmallscaleFluxEmergence} and~\citet{Orozco:FluxEmergenceInQuietSun} studied flux emergence in a quiet Sun environment,~\citet{Ishikawa:TransientHorizontalFields} and~\citet{Otsuji:FluxEmergence} examined flux emergence in a plage region and near sunspot penumbrae, respectively. Despite the different regimes, all three studies point to the importance of granular convective flows on the properties of emerging flux, as was previously reported by~\citet*{DePontieu:Small-scaleEmergingFlux} using filtergram observations with the Swedish Solar Telescope (SST).

In tandem with the improved quality of observational data available, numerical simulations of magnetic flux emergence have, in recent years, become increasing sophisticated in terms of the physics included in the models. For instance,~\citet*{Cheung:FluxEmergenceInGranularConvection} performed 3D radiative MHD simulations to model the rise of buoyant magnetic flux tubes through the near-surface layers of the convection zone and the overlying photosphere. The observational diagnostics of flux emergence in granular convection from these models compare favorably with the aforementioned studies performed with SOT data~\citep[e.g.][]{Centeno:SmallscaleFluxEmergence}.

More recently, several groups have begun to model self-consistently the magnetic connection between the near-surface convection zone layers all to way to the corona. For instance, Tortosa et al. (in prep.) extended the simulations of~\citet*{Cheung:FluxEmergenceInGranularConvection} to include the evolution of the emerging magnetic field in the overlying chromosphere and transition region.~\citet*{MartinezSykora:TwistedFluxEmergence} carried simulations of the emergence of twisted flux tubes from the convection zone into the corona, including effects such as thermal conduction along magnetic field lines and non-LTE radiative terms. These numerical studies yield important new insights for behavior of emerging flux in the chromosphere and above. Although the magneto-convection simulations of~\citet*{Abbett:MagneticConnection} and~\citet*{Isobe:ConvectionDrivenEmergence} were not initialized with buoyant magnetic flux tubes, both studies show that convective flows interacting with an ambient field naturally lead to the creation of small-scale magnetic loops~\citep[see also][]{SteinNordlund:SmallScaleMagnetoconvection}, which emerges through granular upflows and interacts with the pre-existing magnetic fields in the atmosphere.

In~\citet{Cheung:FluxEmergenceInGranularConvection}, we presented a study of magnetic flux emergence in granular convection by means of numerical radiative MHD simulations with the MURaM code~\citep{Voegler:MURaM}. The simulations in that study began with a buoyant magnetic flux tube embedded in the near-surface layers of the convection zone. The simulations were restricted to flux tubes with initials longitudinal fluxes up to $10^{19}$ Mx. In this paper, we extend our previous study to flux tubes with a flux on the order of $10^{20}$ Mx, comparable to the flux content of mid-sized ephemeral regions~\citep*{Hagenaar:Ephemeralregions}.

The paper is structured as follows. In section~\ref{sec:setup}, we describe the setup of the numerical experiments. In section~\ref{subsec:subsurface_evolution}, we discuss the subsurface evolution of the rising flux tube. In section~\ref{subsec:observational_diagnostics}, we compare the observational characteristics of our modeled emerging flux regions with observational results with particuar emphasis on Hinode SOT results. Finally, in section~\ref{sec:discussion}, we discuss the implications of the this work for our understanding of emerging flux regions.

\section{Simulation setup}
\label{sec:setup}
We first obtained a relaxed 3D non-magnetic model of the near-surface layers of the convection zone by means of numerical simulation with the MURaM code, which solves the radiative MHD equations on a Cartesian grid. The simulations take into account the effect of radiative energy exchange and changes in partial ionization in the equation-of-state (EOS). The code has been previously used to investigate various aspects of solar surface magnetism including solar faculae~\citep{Keller:Faculae}, magneto-convection in the quiet Sun, plage regions~\citep{Voegler:Nongrey,Voegler:MURaM,Shelyag:StokesDiagnostics} and mixed polarity regions~\citep{Khomenko:MixedPolarity}, umbral magneto-convection~\citep*{Schuessler:UmbralConvection}, solar pores~\citep{Cameron:Pores}, the photospheric reversed granulation~\citep*{Cheung:ReversedGranulation}, the solar surface dynamo~\citep{Voegler:SolarSurfaceDynamo,Schuessler:StrongHorizontalFields} and, as a predecessor to the present study, magnetic flux emergence in granular convection~\citep*{Cheung:FluxEmergenceInGranularConvection}. All these simulations have been confined to depths below the photospheric base of $1.8$ Mm or less. In order to accommodate for the deeper layers considered here, we use EOS tables from the OPAL project~\citep*{OPAL} for a solar gas mixture with abundances from~\citet*{AndersGrevesse:SolarAbundances}.

The simulation domain has horizontal dimensions $24\times 18$ Mm$^2$ and a height of $5.76$ Mm. The horizontal and vertical grid spacings are $25$ km and $16$ km respectively. The level $z=0$ corresponds to the mean geometrical height where the continuum optical depth at $500$ nm is unity (i.e. $\tau_{500}=1$) and is located approximately $300$ km ($2$ pressure scale heights) below the top boundary. The choice of this location for the top boundary was motivated by the need to keep the time step sufficiently large ($\Delta t$ is as low as $0.01$ s due to the high Alfv\'en speeds in strong field regions) for the simulations to progress. Above the top boundary, the magnetic field is matched to a potential field configuration at each time step. Periodic boundary conditions are imposed at the side boundaries and open boundary conditions allowing for smooth in-/out-flow are imposed at the top and bottom boundaries. For details, see~\citet{Cheung:FluxEmergenceInGranularConvection}.

Using the model convection zone/photosphere as the ambient non-magnetic state, an initially horizontal, twisted magnetic flux tube was introduced into the convection zone at a depth of $z=-3.9$ Mm. The longitudinal and transverse components of the magnetic field have the form 
\begin{eqnarray}
B_l(r) & = & B_0 \exp{(-r^2/R_0^2)}, \label{eqn:long_profile}\\
B_\theta(r) & = & \frac{\lambda
r}{R_0}B_l,\label{eqn:transverse_profile}
\end{eqnarray}
where $r\in[0,2R_0]$ is the radial distance from the tube axis and
$R_0$ the characteristic tube radius. For the two (runs A and B) simulations presented in this paper, $R_0=600$ km
and the field strength at the tube axis is $B_0=14$ kG. The longitudinal flux of the tube is  $\Phi_0 = \int B_l {\rm d}S= 0.98\pi R_0^2 B_0 = 1.55\times10^{20}$ Mx. The dimensionless twist parameter specifies the relative strength of the longitudinal and transverse field components. Simulation runs A and B have twist parameters of $\lambda=0.1$ and $\lambda=0.25$ respectively.

Although the initial field strength ($B_0 = 14$ kG) of the tube is several times larger than the strongest fields observed at the solar surface, it is important to note that
the near-surface convection zone and overlying atmosphere are strongly stratified. In fact, the initial depth of the tube is $7.7$ pressure scale heights below the base of the photosphere ($\tau_{500}=1$). At this depth, the plasma-$\beta$ (ratio of gas and magnetic pressures) at the tube axis is $\beta \approx 30$.

To keep the divergence of the total (Maxwell + gas pressure + viscous) stress tensor continuous throughout the domain, the initial thermodynamic state of plasma enclosed within the tube was modified. We imposed a sinusoidal profile for specific entropy distribution within the tube so that  $s=11.9 R^\star$ at $x=0$ and $x=24$ Mm (lowest entropy at side boundaries) and $s=12.1R^\star$ at $x=12$ Mm, where $R^\star$ is the universal gas constant. The prescription of such a distribution for the specific entropy translates into a density perturbation for the tube which is buoyant at $x=12$ Mm and anti-buoyant at the side boundaries. Thus the differential buoyancy along the flux tube induces it to develop into an $\Omega$-loop configuration.

\section{Simulation Results}
\label{sec:results}
In section~\ref{subsec:subsurface_evolution}, we first discuss aspects of the subsurface evolution of the buoyant flux tubes in the simulations. Thereafter, in section~\ref{subsec:observational_diagnostics}, we discuss the photospheric diagnostics of our simulated emerging flux regions and their relation to observations.

\subsection{Subsurface evolution}
\label{subsec:subsurface_evolution}
\subsubsection{Horizontal expansion, deformation and fragmentation}
\label{subsubsec:horizontal_expansion}

\begin{figure}
\centering
\includegraphics[width=0.48\textwidth]{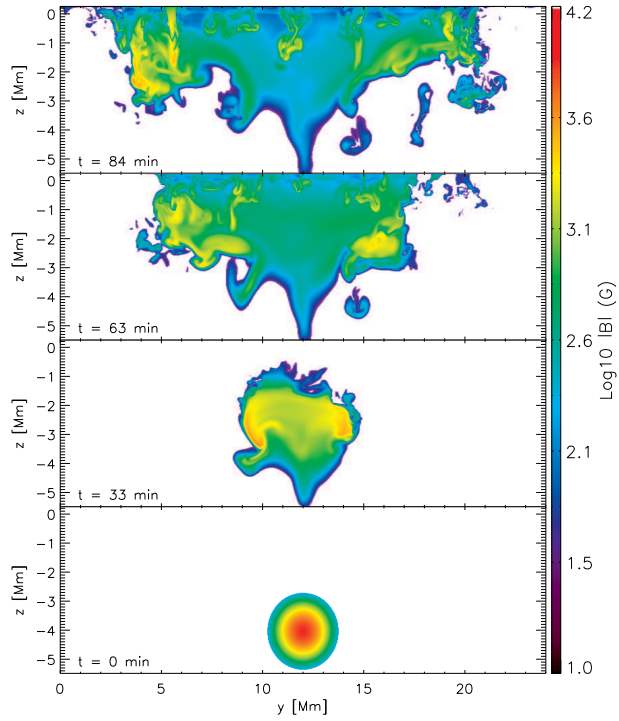}
\caption{Time sequence of vertical cross-sections (at $x=12$ Mm) through the rising magnetic flux tube in Run A (twist parameter $\lambda=0.1$). The rise of the flux tube over multiple pressure scale heights is accompanied by a horizontal expansion which leads to a sheet-like structure below the photosphere.
}\label{fig:cs_plot}
\end{figure}

Figure~\ref{fig:cs_plot} shows a time sequence of cross sections of the tube at $x=12$ Mm in Run A. Between $t=0$ and $t=33$ min, this section of the tube had risen a distance comparable to its initial diameter. Two effects are of particular interest. First of all, the tube tends towards a fragmentation into two counter-rotating vortex rolls (identifiable as magnetic concentrations regions of particularly high field strength) with opposite signs of out-of-plane component of the vorticity. This effect is well-known from earlier numerical studies of the buoyant rise of untwisted and slightly twisted flux tubes and is explained by consideration of the aerodynamic pressure difference across the tube interior and the external flow around it~\citep{Schussler:buoyancyrevisited,Moreno-InsertisEmonet:RiseOfTwistedTubes,EmonetMoreno-Insertis:PhysicsOfTwistedFluxTubes,Fan:2DTubes,Cheung:MovingMagneticFluxTubes}.

In addition, the rise of the tube is accompanied by a strong horizontal expansion, which is expected for the adiabatic rise of a fluid parcel (magnetic or otherwise) over multiple pressure scale heights. By the time it has reached the photosphere, its aspect ratio is such that it appears more like a magnetic sheet than a `typical' flux tube. This result confirms the scenario sketched out in Fig. 4 of~\citet*{SpruitTitleVB:WeakMixedPolarityBackground}, wherein they describe how a flux tube rising over a few pressure scale heights must expand laterally to become a flux sheet. This has potentially important implications for helioseismic attempts to detect emerging flux regions before they erupt onto the solar surface, namely that a flux sheet model is perhaps a more appropriate model than a flux tube model for helioseismic inversions.

\begin{figure}
\includegraphics[width=0.49\textwidth]{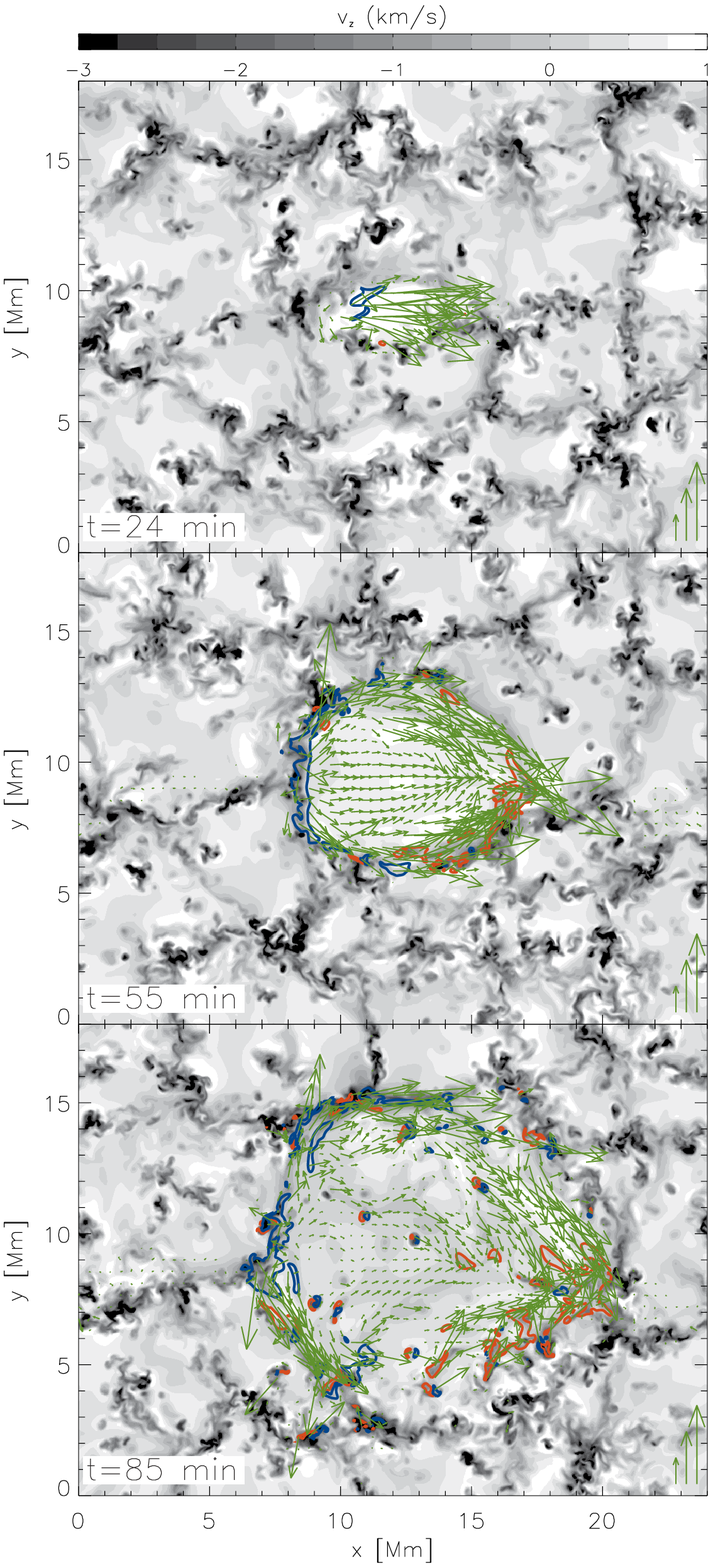}
\caption{Vertical velocity pattern (greyscale) and magnetic field distribution showing the passage of the buoyant magnetic tube through a layer in the convection zone at a depth of $2.3$ Mm. Blue and red contours respectively delineate positive and negative polarity regions with vertical field strength $B_z\ge 1$ kG and the overlaid vectors indicate the direction and strength of the horizontal components of the field. For reference, the three vectors at the lower-right corners of each panel have horizontal strengths of $1,2$ and $3$ kG.}\label{fig:mag_bubble}
\end{figure}

A direct consequence of the strong expansion of the tube during its ascent toward the photosphere is the progressive weakening of the field with decreasing depth (in the absence of variations along the tube axis and under the assumption of flux freezing, $B \propto \varrho$). In contrast to this, mass conservation entails that the amplitude of flow velocities in the convection zone increases toward the surface (the Mach number $\mathcal{M} \sim 0.1$ at the interface between the convection zone and photosphere). These two factors act together in such a way that, with decreasing depth, the influence of the convective flows on the field evolution becomes increasingly important.

Figure~\ref{fig:mag_bubble} shows the passage of the rising tube through a layer of the convection zone at a depth of $2.3$ Mm. The vertical flow field (grey-scale) shows convective cellular patterns not unlike the surface granulation. At this depth, however, the upflow cells have a typical length-scale of $\approx 4-5$ Mm (as opposed to $\approx 1$ Mm  at the surface) and the downflow network already shows signs of fragmentation into separated downflow channels~\citep*[cf.][]{Benson:SupergranularScaleConvection}. At $t=24$ min, the top of the rising tube begins to cross this layer. As the magnetic tube continues to rise, the upflow cell created by the rising tube expands and sweeps aside previously existing downflows. A comparison of the buoyancy and drag (by the downflows) forces experienced by a rising tube shows that for a magnetic tube to rise against downflows~\citep{Parker:Drag,Moreno-Insertis:Risetimes,Fan:3DTubeConvection,Cheung:FluxEmergenceInGranularConvection}, it must have an internal magnetic field strength 
\begin{equation}
B \gtrsim \left(\frac{2C_D \gamma_1}{\pi} \right)^{1/2}\left(\frac{H_p}{R}\right)^{1/2} B_{\rm eq},\label{eqn:force_comparison}
\end{equation}
\noindent where $C_D\sim O(1)$ is the drag coefficient, $\gamma_1$ Chandrasekhar's first adiabatic exponent, $H_p$ the local pressure scale height, $R$ the tube radius, and $B_{\rm eq}$ the equipartition field strength (i.e. the magnetic field strength such that the magnetic energy density is equal to the kinetic energy density of a downflow with speed $v$). At a depth of $2.3$ Mm, typically $90\%$ of the downflowing plasma has a speed of $2$ km s$^{-1}$ or less. Taking $v=2$ km s$^{-1}$ and a mean density of $\langle \varrho \rangle = 1.7\times 10 ^{-5}$ g cm$^{-3}$ at this depth, we have $B_{\rm eq} \approx 3 $ kG. In Fig.~\ref{fig:mag_bubble}, we have seen that the emerging tube is associated with an upflow cell several Mm in diameter. Taking $R = 4 H_p$ (a conservative estimate) and taking $2 C_D \gamma_1 / \pi \approx 1$, criterion (\ref{eqn:force_comparison}) tells us that the emerging magnetic field must have a field strength $B\gtrsim 1.5$ kG. The flux tube in our simulation exhibits field strengths of $2-3$ kG at this depth. In accordance with the above estimate, it is able to overcome the convective downdrafts and emerge.

\subsubsection{Effect of twist on the rate of flux emergence}

\begin{figure}
\centering
\includegraphics[width=0.49\textwidth]{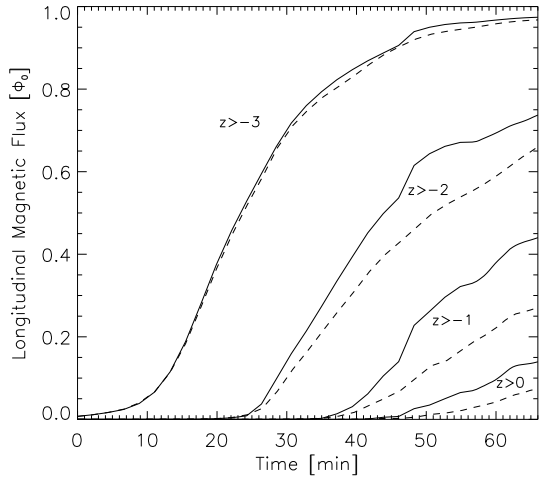}
\caption{Longitudinal magnetic flux crossing the $x=12$ Mm vertical plane. The dashed and solid lines indicate values from the runs A ($\lambda=0.1$) and B ($\lambda = 0.25$) respectively. The curves labeled $z>-3$ indicates the amount of longitudinal flux above a depth of $3$ Mm etc. The amount of longitudinal flux in the photosphere is indicated by curves labeled $z>0$. Fluxes are given in units of the initial longitudinal flux of the tube, $\Phi_0 = 1.55\times10^{20}$ Mx.}\label{fig:fluxes}
\end{figure}

Figure~\ref{fig:fluxes} illustrates the rise of the magnetic flux by displaying plots of the longitudinal magnetic flux crossing the vertical plane $x=12$ Mm as functions of time for the two simulation runs (i.e. $\int B_x dydz$). Curves labeled $z>-3$ indicate the amount of longitudinal flux above a depth of $3$ Mm (the flux tube initially resides just below this depth), which is almost the same for the two simulation runs. Since the initial buoyancy of the flux tubes is the same (to first order in twist parameter $\lambda$),  this is not a surprising result. Toward shallower layers, the results from the two simulation runs diverge from each other systematically such that the rate of transport of longitudinal flux is always higher in the more twisted case (run B). In particular, the rate of longitudinal flux emerging from the convection zone into the photosphere ($z>0$) in run B is almost double that of run A ($\sim 10^{18}$ Mx/min as opposed to $\sim 5\times 10^{17}$ Mx/min).

This finding is consistent with the results of~\citet*{Murray:NonconstantTwist}. By means of MHD simulations of the buoyant rise of magnetic flux tubes through an idealized stratified layer mimicking the convection zone and overlying atmosphere, they found that magnetic tubes with higher degrees of twist (and therefore greater magnetic tension) have higher rates of emergence into the atmosphere. In their case, they can cast the results in the context of a magnetic buoyancy instability~\citep{Acheson:InstabiltyByMagneticBuoyancy,Archontis:EmergenceIntoCorona,Murray:EmergenceParameterStudy}, we cannot apply the same criterion to our simulations because it hinges on the assumption of adiabatic perturbations from an initially stationary equilibrium. These assumptions, in the presence of large entropy changes near the photosphere and ambient convective motion, are clearly violated during the evolution of the magnetic structure in our models.

\subsubsection{Coherent subsurface roots and serpentine field lines of an emerging flux region}
\label{subsubsec:subsurface_roots}

\begin{figure*}
\centering
\includegraphics[width=\textwidth]{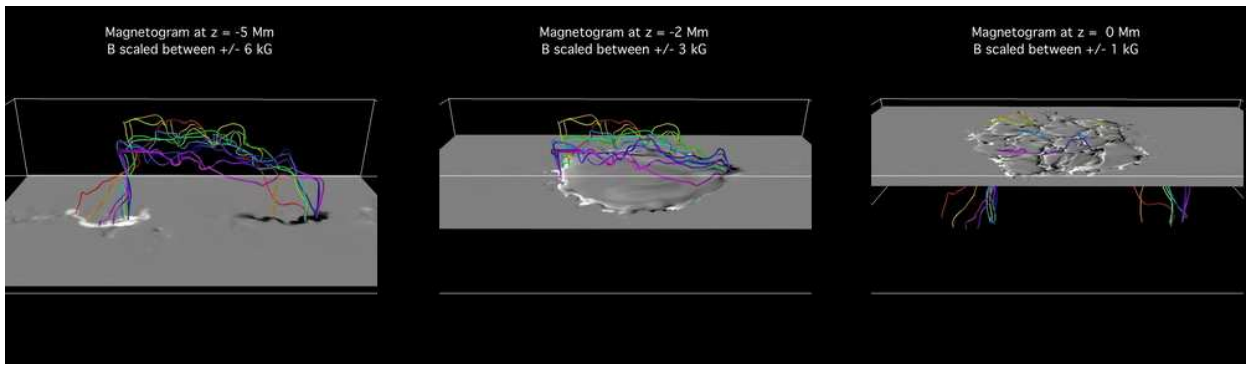}
\caption{The magnetic configuration of the emerging flux region from its subsurface roots to the photosphere (run A, $\lambda=0.1$) at $t=51$ min. The synthetic magnetograms (grey-shading of $B_z$ on horizontal planes) at different depths show the transition from a coherent bipole in the subsurface roots to the complex mixed-polarity pattern at the surface. The serpentine appearance of the field lines near the surface highlight the effect of granular dynamics on the morphology of emerging flux. This figure is also available as an MPEG animation.}\label{fig:subsurface_roots}
\end{figure*}

Figure~\ref{fig:subsurface_roots} shows three synthetic magnetograms ($z=-5$ Mm, $z=-2$ Mm and $z=0$ Mm) from the emerging flux region in Run A ($\lambda=0.1$) at $t=62$ min. Also shown are magnetic field lines. Deep down at $z=-5$ Mm, the field lines group together as rather coherent bundles to give a tidy bipolar structure in the magnetogram. Following the field lines higher up near the surface, the horizontal expansion of the rising tube separates the field lines and the granular dynamics undulates them to form a set of serpentine field lines. This result is consistent with the detailed observations of emerging flux regions by~\citet*{StrousZwaan:SmallScaleStructure} and~\citet{Pariat:ResistiveEmergence}, who report that magnetic field lines in emerging flux regions emerge already undulated. Our result naturally identifies convective dynamics as the cause of the serpentine structure of field lines.

\begin{figure}
\centering
\includegraphics[width=0.45\textwidth]{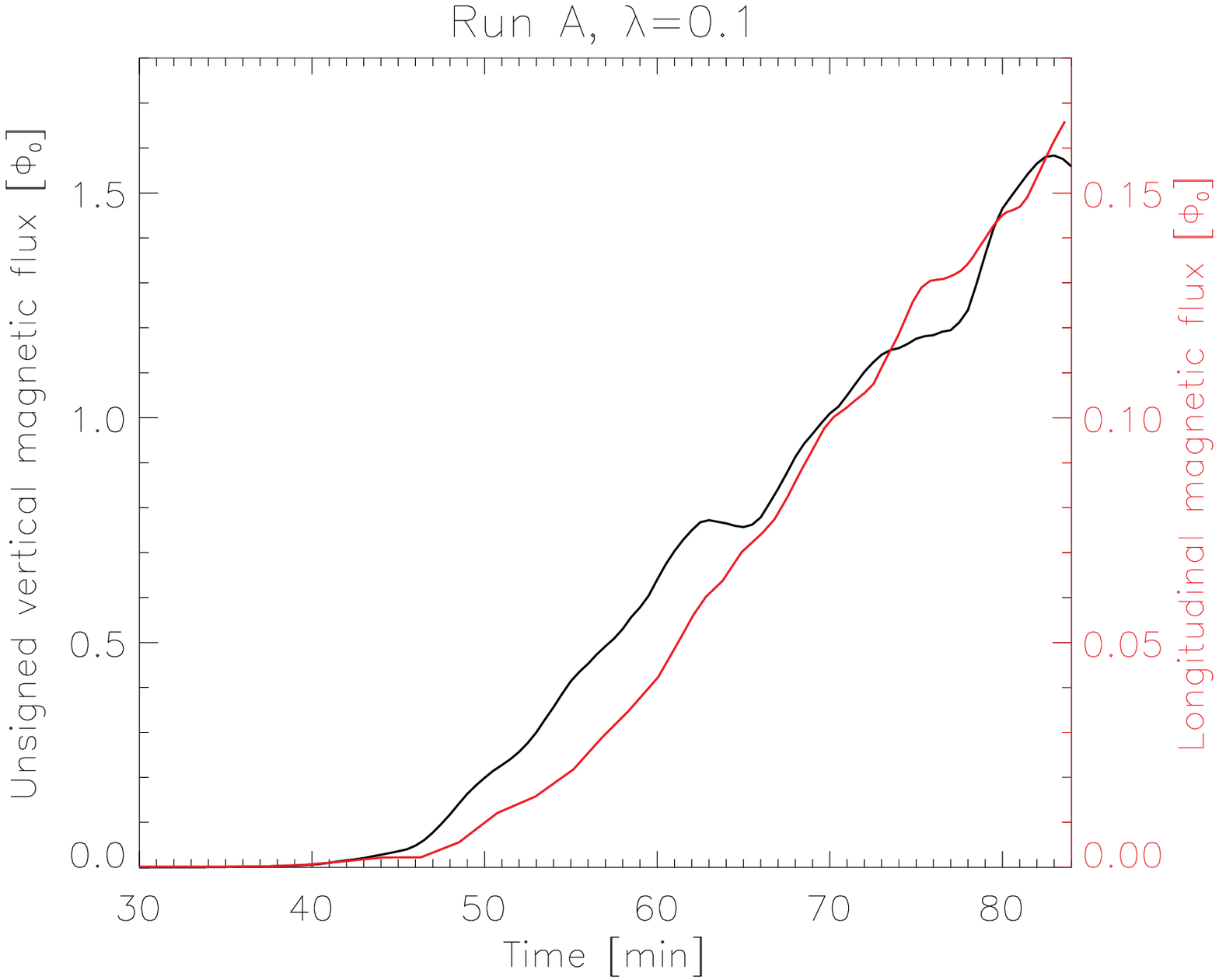}
\caption{Comparison between the longitudinal flux emerged into the photosphere ($\Phi_{\rm long} = \int_{z\ge 0} B_x dydz$) and the corresponding unsigned vertical flux at the photospheric base ($\Phi_{\rm unsigned} = \int |B_z(z=0)| dxdy$). 
}\label{fig:compare_unsigned}
\end{figure}

For a simple $\Omega$-loop emerging through the surface, the ratio of the total unsigned magnetic flux at the surface ($\Phi_{\rm unsigned} = \int_{z=0}|B_z|dxdy$) and the longitudinal flux that has actually emerged ($\Phi_{\rm long}= \int_{z\ge 0} B_x dydz$, measured through a vertical plane normal to the tube axis) is $\Phi_{\rm unsigned}/\Phi_{\rm long}=2$. In the case of flux emergence with serpentine field lines, this ratio is higher. Fig.~\ref{fig:compare_unsigned} shows time plots of both the longitudinal flux crossing $x=12 $ Mm above the convection zone (red line) and the total unsigned flux crossing the base of the photosphere (black line) for Run A ($\lambda=0.1$). Throughout the period of emergence activity, $\Phi_{\rm unsigned}/\Phi_{\rm long}\approx 10$, which indicates that emerging field lines are,  on average, undulated $5$ times by the convective flow. Although the flux emergence rate in Run B ($\lambda=0.25$) is higher, the ratio of the two fluxes at any given time is also $\approx 10$. This common value of $\Phi_{\rm unsigned}/\Phi_{\rm long}$ between the two runs is attributed to the fact that the spatial extents of the two emerging flux regions at the surface are similar.

\subsubsection{Vertical transport of magnetic helicity}
\label{subsubec:magnetic_helicity}
\begin{figure}
\includegraphics[width=0.5\textwidth]{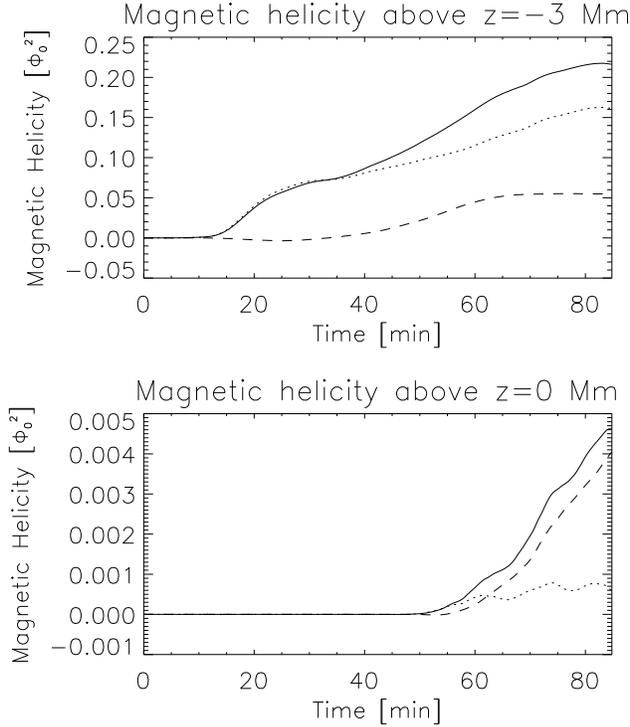}
\caption{Cumulative magnetic helicity injected above $z=-3$ Mm (upper panel) and $z=0$ (lower panel) for simulation Run A (twist parameter $\lambda=0.1$). Helicity injection by the emergence term is indicated by dotted lines, the shearing term is indicated by dashed lines and their sum is indicated by solid lines.}\label{fig:helicity}
\end{figure}
In the limit of ideal MHD, the injection of magnetic helicity into an infinite half space (i.e. $\{(x,y,z) | z \ge z_0 \}$) from the bottom boundary ($z=z_0$) is described by the Poynting theorem~\citep*{Berger:MagneticHelicity}:
\begin{equation}
\frac{dH}{dt} = 2 \oint [ (\vec{A}_p \cdot \vec{B}) v_z  - (\vec{A}_p \cdot \vec{v}) B_z  ]dxdy,\label{eqn:helicity_flux}
\end{equation}
\noindent where $\vec{v}$ is the plasma velocity and $\vec{A}_p$ is a unique vector potential satisfying the boundary and gauge conditions
\begin{eqnarray}
\hat{z}\cdot (\nabla \times \vec{A}_p) & = & B_z,\\
\hat{z}\cdot \vec{A}_p & = & 0 ,\\
\nabla \cdot \vec{A}_p & = & 0.
\end{eqnarray}
\noindent where $\hat{z}$ is the unit vector in the $z$-direction. The first term in Eq. (\ref{eqn:helicity_flux}) describes the bodily transport of twisted magnetic structures by vertical flows through the base plane (emergence term). The second term describes the transport of helicity due to the braiding of magnetic field lines by horizontal shearing motion at the boundary (shearing term).

In accordance with our periodic side boundary conditions, we use Fourier transforms~\citep*[cf. ][]{Chae:HelicityFlux} to calculate $\vec{A}_p$ for a given $B_z$ distribution. The~\emph{cumulative} magnetic helicity injected above a depth of $3$ Mm and above the base of the photosphere ($z=0$) are plotted in the upper and lower panels of Fig.~\ref{fig:helicity}, respectively. The values of helicity are normalized to $\Phi_0^2$ (square of the original longitudinal magnetic flux of the tube).

We find that within the convection zone, at $z=-3$ Mm, the bulk of the vertical helicity transport is due to the emergence term, whereas near the surface, at $z=0$, shearing term dominates to carry most of the helicity transport. Again, the reason for this change can be put into the context of the horizontal expansion experienced by the tube. In deeper layers below the surface, the ratio of the tube radius to the pressure scale height is small, so that as parts of the tube traverses the $z=-3$ Mm plane, it experiences relatively little expansion. As we found in section~\ref{subsubsec:horizontal_expansion}, the diminishing pressure scale heights near the top of the convection zone causes a dramatic expansion of the tube. In turn, the horizontal flows associated with this expansion enhance the shearing contribution to helicity transport.

In light of the work by Manchester et al.~(\citeyear{Manchester:EruptionOfEmergingFluxRope}) and~\citet*{Magara:EmergingFluxSurfaceFlows}, who reported that the expansion of rising twisted magnetic tubes into the solar atmosphere drives systematic shear flows, it is expected that the shearing term becomes increasingly important towards the surface. In fact,~\citet*{Magara:HelicityInjection} finds from their emerging flux simulations that although the shearing term is initially smaller than the emergence term (for the photospheric base), the shearing term persists longer and becomes the dominant contributor to helicity injection into the solar atmosphere.

Strictly speaking, Eq.~(\ref{eqn:helicity_flux}) applies to an infinite half-space with no side boundaries. So the use of Fourier transforms for the computation of $\vec{A}_p$ is only consistent with the scenario of identical emerging flux regions periodically spaced apart. To mimick the scenario of a single emerging flux region embedded in field-free surroundings, one can numerically evaluate the Biot-Savart integral for $\vec{A}_p$ over the lower boundary of an infinite half-space by setting $\vec{B}=0$ outside the simulation domain~\citep*[e.g.][]{Berger:MagneticHelicity,Berger:HelicityInSpacePhysics,Welsch:MutualHelicity,Pariat:HelicitFluxDensity,Jeong:HelicityInjection}. Application of the two methods for this study show that the Fourier transform method typically yields helicity fluxes (and hence cumulative injected helicities) which are $\sim 10-20\%$ higher than the direct integration method.~\citet*{Jeong:HelicityInjection} also compared helicity fluxes from both methods and found a comparable excess from the Fourier transform method.

\subsection{Observational diagnostics: comparison between model and Hinode Solar Optical Telescope observations}
\label{subsec:observational_diagnostics}

\subsubsection{A magnetic inversion layer above granular upflows}
\label{subsubsec:inversion_layer}
\begin{figure*}
\centering
\includegraphics[width=0.7\textwidth]{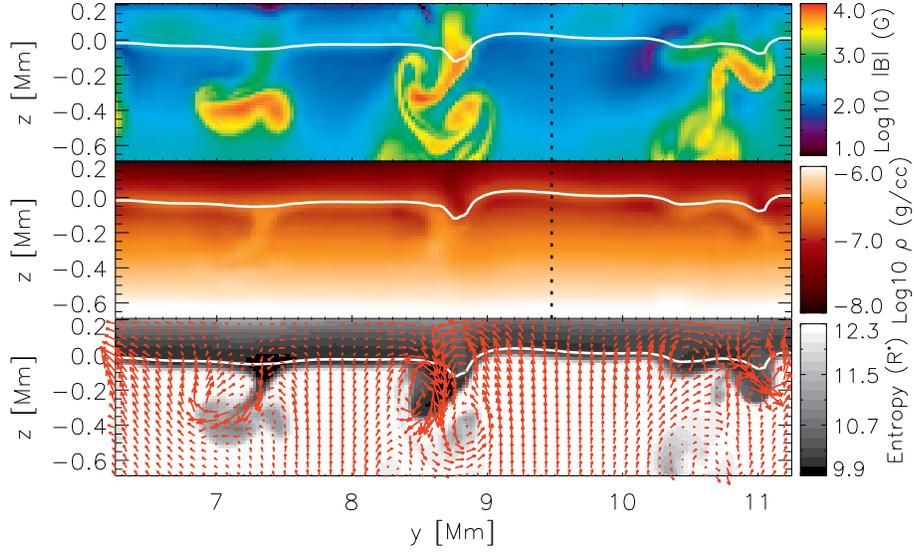}
\caption{Vertical cross-sections (at $x=12$ Mm) of magnetic field strength $|B|$, mass density $\varrho$ and specific entropy $s$ (top to bottom, respectively) at in the near-surface layers of the emerging flux region in Run A (twist parameter $\lambda=0.1$). In all three panels, the white solid line near $z=0$ indicates the $\tau_{500}=1$ surface. In the bottom panel, the red arrows indicate the fluid velocity in the plane. Immediately above this surface is a~\emph{magnetic inversion layer}. Vertical profiles of $|B|$ and $\varrho$ along the black dashed line are shown in Fig.~\ref{fig:inversion_line}.}\label{fig:inversion_layer}
\end{figure*}

\begin{figure}
\centering
\includegraphics[width=0.48\textwidth]{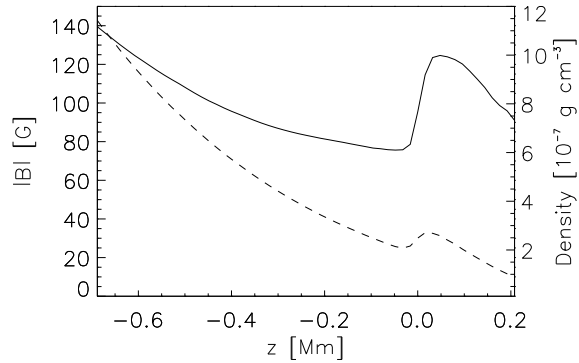}
\caption{Vertical profiles of $|B|$ (solid line) and $\varrho$ (dashed line) taken along the vertical dashed line in Fig.~\ref{fig:inversion_layer} clearly indicate the presence of the magentic inversion layer.}\label{fig:inversion_line}
\end{figure}

During the period of intense emergence activity, plasma both above and below the $\tau_{\rm 500}=1$ surface is permeated by magnetic field. This is shown by the cross-sectional plots in Fig.~\ref{fig:inversion_layer}. In the top panel, we find that below $\tau_{\rm 500}=1 $ (in granular upflows), the magnetic field strength generally increases with depth. This is consistent with the quasi-adiabatic expansion of frozen-in magnetic field in expanding upflows. Immediately above the $\tau_{500}=1$ surface on top of granules, however, is a layer where the magnetic field strength exceeds that of the plasma below the surface. This~\emph{magnetic inversion layer} is coincident with an inversion layer in the mass density (central panel of Fig.~\ref{fig:inversion_layer}) where material just above $\tau_{500}=1$ is top-heavy (see Fig.~\ref{fig:inversion_line} for a vertical profiles of $|B|$ and $\varrho$ through a typical upflowing region). The relative increase of $|B|$ across the boundary layer ($\sim 50\%$) is larger than the relative increase of $\varrho$ ($\sim 30\%$).

In non-magnetic solar convection simulations, such a density inversion layer above granules is well-known (see Fig.~18~of~\citet*{SteinNordlund:SolarGranulation} and Fig.~8~of~\citet*{Cheung:ReversedGranulation} for plots of the density history of fluid parcels emerging into the photosphere). The high density layer results from the abrupt radiative cooling experienced by plasma emerging onto the photosphere (see the sharp drop in specific entropy in the lower panel).

By combining the mass continuity equation and the Lagrangian form of the ideal MHD induction equation,
\begin{equation}
\frac{D\vec{B}}{Dt} = -\vec{B}(\nabla \cdot \vec{v}) + (\vec{B}\cdot \nabla) \vec{v},\label{eqn:induction_lagrangian}
\end{equation}
\noindent we can derive Walen's equation for $\vec{B}/\varrho$:
\begin{equation}
\frac{D}{Dt}\left( \frac{\vec{B}}{\varrho}\right) =\left( \frac{\vec{B}}{\varrho} \cdot \nabla \right) \vec{v}.
\end{equation}
\noindent This shows that the compression of plasma also leads to an intensification of the predominantly horizontal magnetic field emerging onto the solar surface. In the absence of the stretching of field lines, the r.h.s.~of this equation vanishes and we have $\vec{B}/\varrho = $ constant. From Fig.~\ref{fig:inversion_line} however, we see that the ratio $|B|/\varrho$ increases with $z$ (especially pronounced above $z=0$). This shows that the stretching of magnetic field lines by the convective flow 1) reduces the weakening of the field due to expansion in the convection zone and 2) leads to the stronger relative change of $|B|$ across the inversion boundary layer in comparison to that of $\varrho$.

\subsubsection{Complex patterns in the surface magnetic field distribution}

\begin{figure*}
\centering
\includegraphics[width=0.9\textwidth]{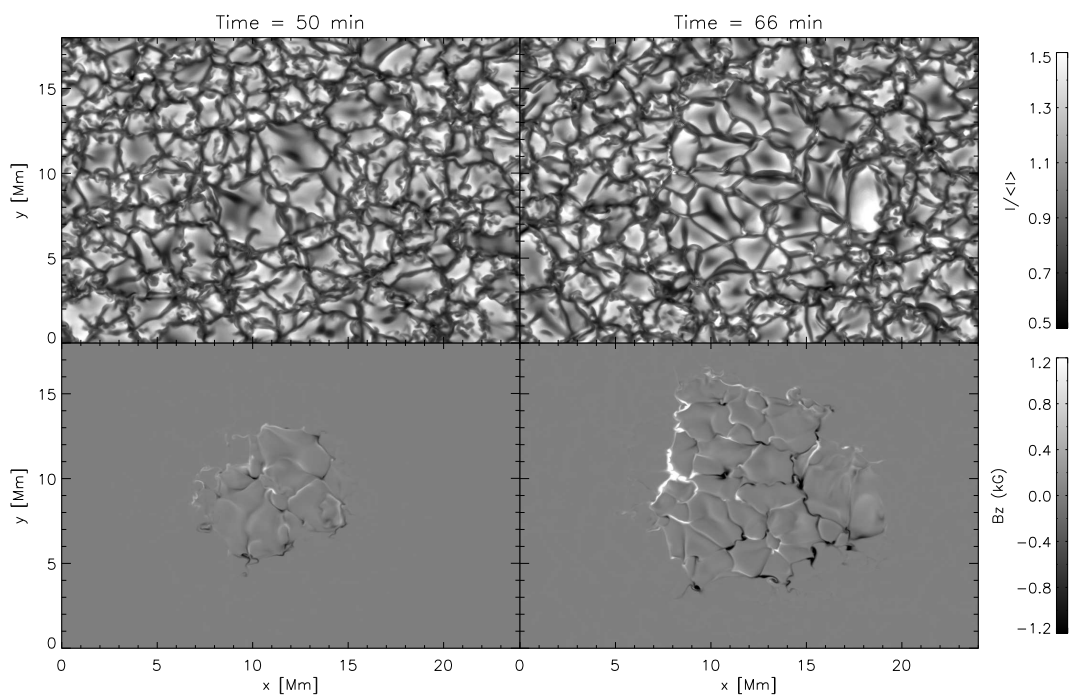}
\caption{Continuum intensity images (at $500$ nm) and synthetic magnetograms of the simulated emerging flux region in Run B (twist parameter $\lambda=0.25$). The vertical component of the magnetic field ($B_z$) is sampled from the $\tau_{500}=0.1$ surface along vertical lines-of-sight. The interaction of the flux tube with granular convective motions is clearly imprinted onto the surface flux pattern, which is far from a simple, tidy pair of opposite polarity flux concentrations. Magnetic bright points in the intergranular lanes can be discerned in the snapshot at $t=66$ min. This figure is also available as an MPEG animation.}\label{fig:mixed_polarity}
\end{figure*}

One of the most striking features of the surface magnetic flux distribution in our simulations is the complexity of the mixed-polarity patterns (see Fig.~\ref{fig:mixed_polarity}). Instead of a pair of coherent flux concentrations of opposite polarity, the simulated emerging flux region (EFR) consists of a large number of flux concentrations residing in the intergranular network of downflows. Although there is a net bipolarity of the flux pattern (there are more positive then negative polarity concentrations on the r.h.s. and vice versa), there is also a considerable amount of mixed-polarity in the interior of the emerging flux region. This is an observational consequence of the undulation experienced by field lines rising against the convective downdrafts (see section~\ref{subsubsec:subsurface_roots}). The net bipolarity of the region becomes more apparent with sufficient spatial averaging (lower resolution).

Although the mixed-polarity pattern is complex, the serpentine nature of the field lines gives it a certain order, namely that the positive polarity concentrations appear at the left edge of granules and stream leftwards, whereas negative polarity concentrations appear at the right edge and stream rightwards. The apparent horizontal motion of individual magnetic elements from the simulated EFRs have speed of about $1-2$ km s$^{-1}$, with magnetic elements near the fringe migrating at a faster pace than those close to the center of EFR. This is consistent with previous reports of moving magnetic features in EFRs~\citep{StrousZwaan:HorizontalDynamics,StrousZwaan:SmallScaleStructure,Bernasconi:EmergingFlux}. In such a scenario where opposite polarity flux concentrations are counter-streaming, the encounter of opposite polarity fields is common. The effects of flux cancellation within an emerging flux region are discussed in the following section.

\subsubsection{Surface flux cancellation as a source of supersonic downflows}
\label{subsubsec:supersonic_downflows}
\begin{figure}
\includegraphics[width=0.47\textwidth]{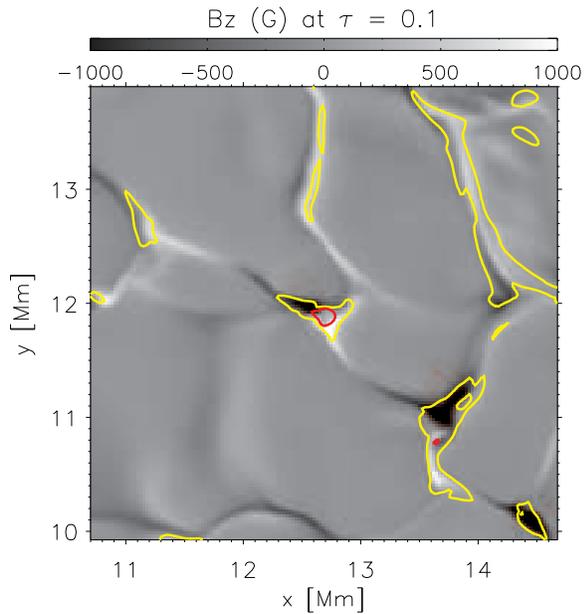}
\caption{Synthetic magnetogram ($B_z$ sampled at $\tau_{500}=1$) of the interior of the simulated emerging flux region (Run B, $\lambda = 0.25$) at $t=60$ min. Yellow and red contours show downflow regions with vertical speeds of $2$ km s$^{-1}$ (yellow contours) and $6$ km s$^{-1}$ (red contours) respectively. This snapshot shows an example of a supersonic downflow at a flux cancellation site. A 3D rendering of this region is shown in Fig.~\ref{fig:reconnection_13600}.}\label{fig:flux_cancellation}
\end{figure}

Encounters between flux concentrations of opposite polarities within an EFR have a number of interesting observational consequences. Firstly, magnetic diffusion between opposite polarity fields and the retraction of inverted U-loops leads to a decrease of the total unsigned flux at the photosphere. Secondly, such flux removal sites may be locations of particularly strong, in some cases even supersonic, downflows. Fig.~\ref{fig:flux_cancellation} shows a small region within the simulated EFR in Run B ($\lambda = 0.25$). Between the pair of opposite polarity flux concentrations in the center of the field of view, we find a vertical downflow with a speed exceeding $6$ km s$^{-1}$ (red solid). In the very center of this downflow, the speed reaches values of up to $10$ km s$^{-1}$. While such a downflow is supersonic ($\mathcal{M}_s \approx 1.3$), it is still sub-Alfv\'enic ($\mathcal{M}_{\rm A} < 1$). This suggests that Maxwell stresses in the magnetic field are primarily responsible for the acceleration of the supersonic downflow.

\begin{figure}
\includegraphics[width=0.34\textwidth,angle=270]{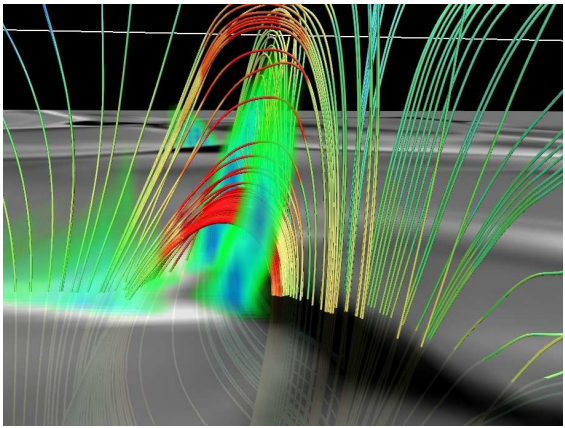}
\caption{3D rendering of the flux cancellation site depicted in Fig.~\ref{fig:flux_cancellation}. The magnetic field lines are colored according to the local value of $v_z$, with red indicating a vertical velocity $v_z < -6$ km s$^{-1}$. The turquoise colored sheets are regions of enhanced $\vec{j}^2/\varrho$ (i.e. sites of enhanced magnetic gradients).}\label{fig:reconnection_13600}
\end{figure}

Figure~\ref{fig:reconnection_13600} shows a 3D rendering of the flux cancellation site depicted in Fig.~\ref{fig:flux_cancellation}. The semi-transparent grey-scale surface shows the footpoint of the opposite polarity magnetic concentrations ($B_z$ at $z=0$) at the photospheric base. Magnetic field lines through the flux concentrations are color-coded by the local values of $v_z$, with reddish colors indicating downflows with $v_z < -6$ km s$^{-1}$. The turquoise colored sheets are region of enhanced $\vec{j}^2/\varrho$ (we divide by $\varrho$ to fold out the effects of density stratification). Clearly, the regions of particularly strong downflows are located at the places where the curvature of the field lines, and hence the magnetic tension, is strongest. A quantitative examination of the forces in the downflow region reveals that: (1) the downwards directed gravitational acceleration ($\varrho\vec{g}$) of the plasma is roughly balanced by the upwards directed gas pressure gradient force ($-
\nabla p_{\rm gas}$) and (2) the vertical component of the Lorentz force ($c^{-1}\vec{j}\times\vec{B}$) is directed downwards and has an amplitude several times that of both the pressure gradient and gravitational forces. Thus the Lorentz force is clearly identified as the cause of supersonic downflows at flux cancellations sites in the photosphere.

\begin{figure*}
\centering
\includegraphics[width=0.9\textwidth]{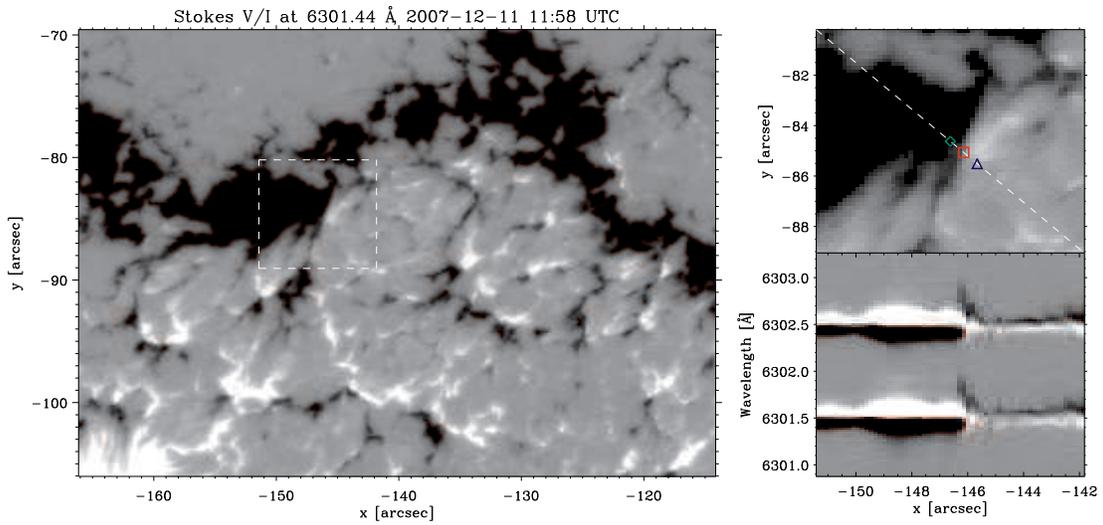}
\caption{Hinode SOT observation of the emerging flux region leading to the formation of AR 10978. The left panel shows an SpectroPolarimeter (SP) scan of the EFR (sampled in the blue wing of the Fe 6301.5 line). The top right panel shows a zoom view of the flux cancellation region. The lower right panel shows the Stokes $V/I$ profiles along the dashed diagonal line in the overlying panel. The redshift of the profiles associated with a supersonic downflow can clearly be seen at the position of the magnetic polarity reversal.
}
\label{fig:sot_cancellation_site}
\end{figure*}

Detections of supersonic downflows ($v_z < -6$ km s$^{-1}$) in the solar photosphere have previously been reported inside a delta sunspot pair~\citep{Pillet:SupersonicDownflowsInDeltaSunspot} and in the penumbra~\citep*{DelToroIniesta:SupersonicEvershedDownflows} of an individual sunspot. More recently, using Hinode SP data,~\citet{Shimizu:HighSpeedMassDownflows} reported that such supersonic downflows in fact appear frequently in a wide range of environments in and outside of sunspots and active regions. They sketched a number of scenarios which may cause such supersonic downflows, one of which involves magnetic reconnection.

\begin{figure}
\centering
\includegraphics[width=0.4\textwidth]{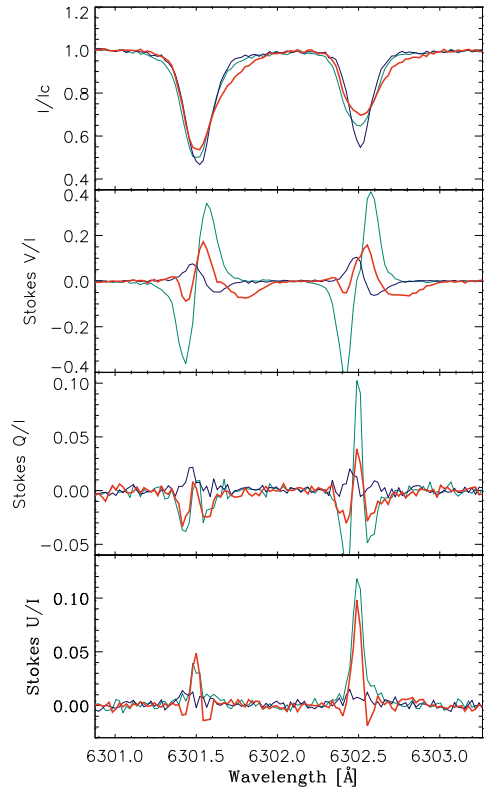}
\caption{Stokes profiles sampled at the three positions shown in Fig.~\ref{fig:sot_cancellation_site}. The red lines indicate profiles at the
polarity inversion line (red square symbol) and the blue and green lines indicate profiles from opposite polarity regions on either side of the inversion line.}
\label{fig:sot_stokes}
\end{figure}

In Fig.~\ref{fig:sot_cancellation_site}, we show an example of a supersonic downflow which is most probably a result of reconnection in or above the solar photosphere. The left panel of this figure shows a Hinode SP scan of the EFR associated with AR 10978. The top right panel shows a zoom view of a small region where there is a clear reversal of the longitudinal magnetic polarity. In the panel immediately below, we show the Stokes $V/I$ profiles sampled along a line indicated by the diagonal white dashed line. At the polarity inversion line, there exists strongly redshifted Stokes $V$ signal. To examine this downflow in further detail, we refer to the plots of the individual Stokes profiles in Fig.~\ref{fig:sot_stokes}. The spatial locations at which the Stokes profiles were sampled are indicated by the diamond (negative polarity), square (polarity inversion line) and triangle (positive polarity) symbols. One of the interesting features of the profiles at the polarity inversion (red curves) is the W-shape of the Stokes $V$ signal, which is consistent with the fact that the pixel is at the polarity inversion line and thus its profiles have contributions from opposite polarity fields~\citep*{Pillet:SupersonicDownflowsInDeltaSunspot,DelToroIniesta:SupersonicEvershedDownflows}. The presence of significant linear polarization indicates horizontal components of $\vec{B}$. Finally, the strong redshift of the profiles (especially in Stokes $I$ and $V$) indicate plasma with downflows speeds approaching $\sim 10$ km s$^{-1}$. All these features are consistent with the scenario of flux cancellation in our simulation (see. Figs.~\ref{fig:flux_cancellation} and~\ref{fig:reconnection_13600}).

Before proceeding to other observational diagnostics of flux emergence, it is important to clarify that although the magnetic loops shown in Fig.~\ref{fig:reconnection_13600} are the result of reconnection between field lines originally pertaining to distinct flux concentrations, the simulations we have performed here do not allow for a detailed study of the reconnection process itself. Due to the location of the top boundary ($z = 300$ km) above the photospheric base, the connectivity of the photospheric field over length scales $l\gtrsim 300$ km is determined by the potential field boundary condition. Consider two opposite polarity magnetic concentrations in the photosphere whose field are originally not mutually connected. As the separation of the flux concentrations decreases, their fluxes are likely to become connected by virtue of the potential field condition even before they are so close that reconnection within the domain takes place to connect the two flux concentrations. In simulations of resistive flux emergence that do include (albeit in an idealized fashion) layers above the photosphere~\citep*{Isobe:EllermanBombs}, the reconnection process is modeled and tends to occur in the chromospheric layers. Both scenarios, however, give a similar end result: namely a set of inverted U-loops which represent new magnetic connections between the flux concentrations. It is the subsequent relaxation of these inverted U-loops, and not necessarily how they were formed, which is of primary concern as to the source of the supersonic downflows. Furthermore, an examination of the terms in the induction equation ($-v_z \partial B_z/\partial z$ vs. magnetic diffusion term) for the scenario in Fig.~\ref{fig:flux_cancellation} reveals that flux removal is predominantly a result of the retraction of inverted U-loops.

\subsubsection{Convective collapse and the formation of kilogauss field and associated bright points}
\label{subsubsec:convective_collapse}

Since convective collapse was first suggested as a possible mechanism for the intensification of photospheric flux tubes to kG strengths~\citep*{Parker:ConvectiveCollapse,Webb:ConvectiveInstability,SpruitZweibel:ConvectiveInstability}, a large body of observational and theoretical work has been carried out to test and to refine the original theory. A particularly important refinement has been the consideration of the inhibitive effects of lateral radiative energy exchange between a flux tube and its surroundings~\citep*{Venkatakrishnan:InhibitionOfConvectiveCollapse}, which essentially violates the adiabaticity condition usually assumed in the convective collapse process. Calculations by~\citet*{Rajaguru:ConvectiveCollapse} of convective collapse in axisymmetric flux tubes show that radiative energy exchange does indeed inhibit convective collapse in flux tubes with $\Phi \lesssim 10^{18}$ Mx. This result is consistent with the observationally determined flux-field strength relation by~\citet{Solanki:IR_12_CC_inhibition}.

The inhibitive effect of lateral radiative energy exchange on field intensification has also been studied in multi-dimensional numerical models of solar surface magneto-convection. In such dynamic MHD models, the magnetic flux concentrations never really begin from equilibrium configurations and the collapse mechanism is often referred to as~\emph{convective intensification}~\citep{Schuessler:TheoreticalAspects,Grossmann-Doerth:ConvectiveIntensification,Steiner:ConvectiveCollapse}. In~\citet*{Cheung:FluxEmergenceInGranularConvection}, we examined the flux-field strength relation for our flux emergence simulation and found - in agreement with~\citet*{Rajaguru:ConvectiveCollapse} - that convective intensification is indeed inhibited by lateral radiative energy exchange for flux concentrations with $\Phi \lesssim 10^{18}$ Mx.

\begin{figure*}
\centering
\includegraphics[width=0.75\textwidth]{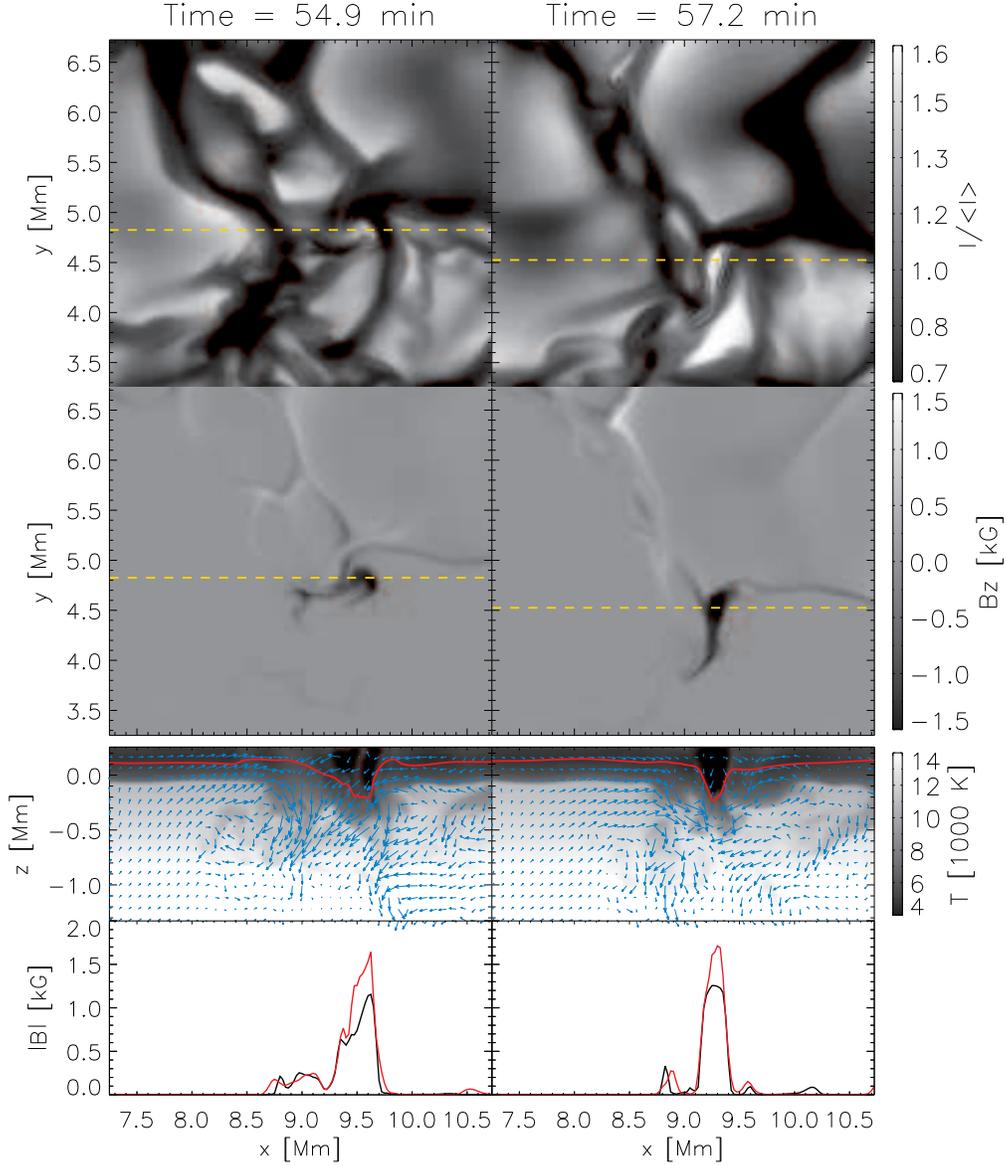}
\caption{Convective intensification as the cause of bright point formation.~\emph{Top row}: continuum intensity (at $500$ nm).~\emph{Second row}: surface synthetic magnetogram
(sampled at the $\tau_{500} = 0.1$ surface).~\emph{Third row}: vertical cuts of the temperature (grey-scale) and velocity field (vectors) along the yellow dashed lines, with the elevation of the $\tau_{\rm 500}=0.1$ surface indicated as red curves.~\emph{Fourth row}: plots of $|B|$ along the yellow dashed line including (red curves, sampled at $\tau_{500}=0.1$) and excluding (black curves, sampled at  $z=0$ km) the effects of the local Wilson depression.} \label{fig:c_collapse}
\end{figure*}

Here we present such an example of a flux concentration undergoing convective intensification from the simulations. Fig.~\ref{fig:c_collapse} shows two snapshots of a magnetic flux concentration resulting from the expulsion of emerge flux to the intergranular lanes in Run A. At $t=54.9$ min, we find a flux concentration trapped in a downflow vertex. In the absence of magnetic fields, such downflow regions are usually relatively dark. However, as the continuum intensity map shows, the flux concentration at this instant already has a brightness comparable to neighboring granules. At constant geometrical height $z=0$ (photospheric base), the field strength is predominantly sub-kG. A couple minutes later, at $t=57.2$ min, convective intensification has proceeded to further strengthen this flux concentration. In this `collapsed' state, even at $z=0$, most of the tube has field strength in excess of $1$ kG. A consequence of the partial evacuation of the flux concentration is the creation of a local Wilson depression, which reaches down to a depth of $z=-300$ km at its center. The presence of such Wilson depressions also means that magnetically sensitive photospheric lines probe deeper into the photosphere (as indicated by the $\tau_{\rm 500}=0.1$ surfaces in Fig.~\ref{fig:c_collapse}, which is close to where the response of the Fe 6301.5 and Fe 6302.5 lines are strongest). So, in addition to a true strengthening of the surface field during convective intensification, the partial evacuation of flux concentrations gives an additional~\emph{apparent} intensification (compare the red and black solid curves in the lower panel of Fig.~\ref{fig:c_collapse}).

At $t=57.2$ min, the magnetic flux concentration appears as a bright point, with continuum intensity $I_{500}/\langle I_{500} \rangle \approx 1.6$. From the Eddington-Barbier relation, we know that for vertical lines of sight, the outgoing intensity $I_\nu$ is given by the source function $S_\nu$ at optical depth unity. So, even though the flux concentration, at $z=0$, is relatively cool compared to surrounding plasma at the same geometrical height, its temperature at $\tau_{500}=1$ is close to $7,000$ K. In contrast, at the geometrical depth of the Wilson depression ($z=-300$ km), the plasma in the adjacent granules can be much hotter ($T\approx 8,000-9,000$ K).~\citet{BellotRubio:PlageFluxTubes} have modeled the structure of plage fields by inversion of observed Stokes profiles assuming a thin flux tube model and reported a similar result. So although the magnetic flux concentration is radiating profusely, its internal temperature (at $\tau_{500}$ = 1) can be maintained above the average value of the quiet Sun by side-wall heating~\citep*{Spruit:FluxTubeEnergyBalance}.

In our simulation, flux concentrations undergoing convective intensification have downflow velocities reaching down to $v_z = -6 $ km s$^{-1}$. Unlike the flux cancellation case described in section~\ref{subsubsec:supersonic_downflows} (where the Lorentz force was responsible for the acceleration to supersonic speeds), the downflows in this scenario of flux concentration is caused by an imbalance of the vertical pressure gradient ($-\nabla p_{\rm gas}$) against gravitation ($\varrho \vec{g}$), sufficient to accelerate the material to $6$ km s$^{-1}$ over one or two minutes. After the formation of the bright point, the magnetic concentration evolves continuously in response to the buffeting of the surrounding granules, with its relative brightness staying above $I/\langle I \rangle > 1 $. This type of evolution resembles the behavior of bright magnetic elements observed by~\citet{Berger:MagneticElements}.

Recently,~\citet{Nagata:ConvectiveCollapse} reported on observations by Hinode SP of a photospheric flux concentration undergoing convective collapse. Using a two component Milne-Eddington inversion, they identified the intensification of the field strength from $500$ G to $2,000$ G. The downflow associated with the collapse grew to $6$ km s$^{-1}$ over a time period of $150$s. A similar scenario has also been inferred by~\citet{BellotRubio:ConvectiveCollapse} from the time evolution of Stokes profiles of infrared lines. All these observed parameters of the collapse process are in good agreement with the example we presented above from our simulation.

\subsubsection{Anomalous transient darkenings}

\begin{figure*}
\centering
\includegraphics[width=\textwidth]{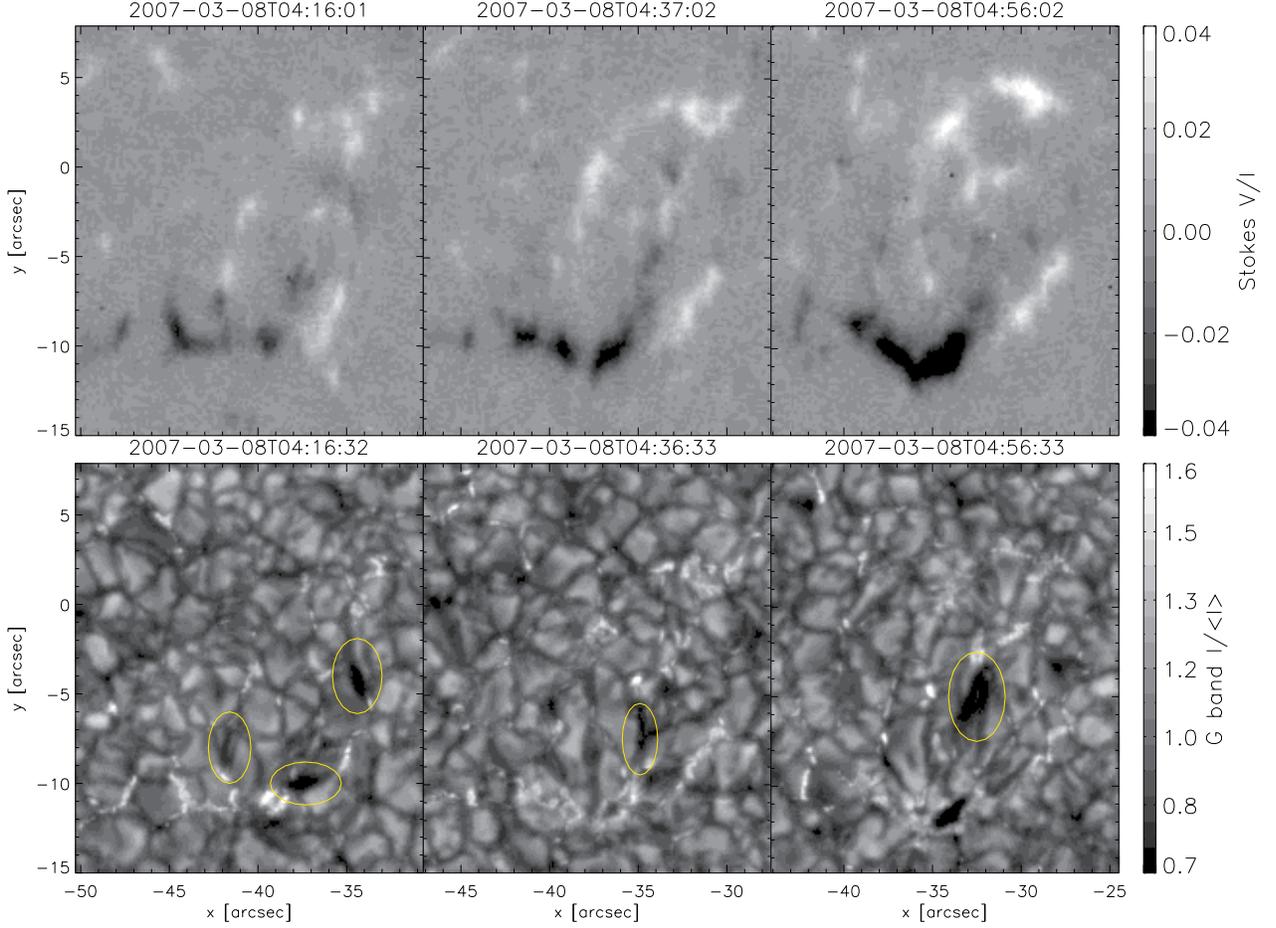}
\caption{Hinode SOT maps of fractional circular polarization (top, NFI Stokes $V/I$) and G-band intensity (bottom, BFI) of an ephemeral region during intense emergence activity. There exists several mixed-polarity patches within the center of the emerging flux region (EFR), an imprint of the granulation dynamics on the emergence process. Observational signatures of flux emergence in the G-band include elongated granules along the direction connecting the two main patches of opposite polarity flux, transient darkenings (marked by yellow ellipses) and bright point formation. }\label{fig:march8ef}
\end{figure*}

The observational study of small-scale flux emergence events in an EFR by~\citet{StrousZwaan:SmallScaleStructure} indicate that transient darkenings (a few Mm in length and with life times of $\approx 10$ min) are associated with the emergence of small-scale flux bundles emerging into the solar atmosphere. Their statistical study showed that the transient darkenings are aligned with upflows on the order of $0.5$ to $1$ km s$^{-1}$ and that the lanes were usually flanked by the formation of bright grains at the ends of the darkening. Based on such diagnostics, they conclude that the transient darkenings correspond to the crests of rising loops, whereas the flanking bright points corresponds to the photospheric footpoint of such loops. Furthermore, they suggest that the `bright grains appear while the magnetic field in the footpoints is concentrating by convective collapse'.

The observations by~\citet*{StrousZwaan:SmallScaleStructure} are, to a large extent, confirmed by the high-resolution observations from Hinode SOT. For instance, Fig.~\ref{fig:march8ef} shows fractional circular polarization  (NFI $6302$~\AA~Stokes $V/I$, top row) and G-band intensity (BFI at $4305$ \AA) maps of an EFR near disk center (the coordinates are in arcsec from disk center). The data was reduced using the FG\_PREP routine in Solarsoft~\citep*{Freeland:Solarsoft} and image alignment was performed by manually comparing the G-band and Stokes I intensity images.

In all three G-band images, one can find darkenings ($2-3$ arcsec long, marked by yellow ellipses) in the EFR which are thicker and darker than typical intergranular lanes (though the dark feature at $(x,y)=(-34'',-12'')$ in the last frame is actually a pore). Upon close inspection, one also finds that bright grains typically appear at the ends of such transient darkenings. Analysis of the dynamics in those regions in the simulation reveals that convective intensification (see section~\ref{subsubsec:convective_collapse}) is indeed the mechanism responsible for the formation of bright grains, thereby confirming the hypothesis of~\citeauthor*{StrousZwaan:SmallScaleStructure}.

As reported by~\citet*{Cheung:FluxEmergenceInGranularConvection}, the rise of small-scale flux tubes, given sufficient twist, are also able to create disturbances in the granulation pattern. For instance, a tube with $10^{19}$ Mx flux and a twist parameter of $\lambda=0.5$ remains sufficiently coherent during its rise through granular convection that its horizontal expansion at the photospheric base modifies the local granulation pattern and leads to the transient appearance of dark upflows. A general disturbance of the granulation pattern in our simulated EFR can also be seen in Fig.~\ref{fig:mixed_polarity}. In the intensity image on the right, we seen that within the EFR, the granular structures tend to be larger and more elongated than the `typical' quiet Sun granule. 

\begin{figure}
\includegraphics[width=0.46\textwidth]{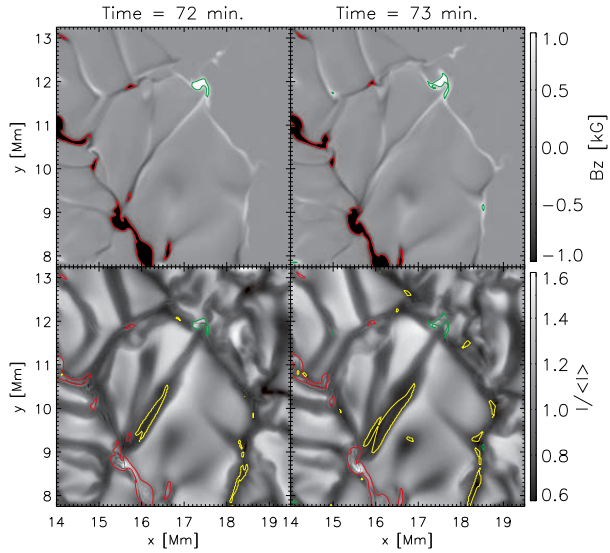}
\caption{Example of a transient darkening in the simulated emerging flux region (Run A, $\lambda=0.1$). The upper and lower panels respectively show the vertical field component (sampled at $\tau_{500}=0.1$) and emergent continuum intensity at $500$ nm. The transient darkening corresponds to the crest of a granular-scale $\Omega$ loop emerging at the photosphere (yellow contours indicate magnetic upflows with $I/\langle I \rangle < 0.7$ and $|B_{\rm hor}|>100$ G and $v_z = 0.5$ km s$^{-1}$). The footpoints of the loop correspond to the opposite polarity flux concentrations at ends of the darkening (the green and red contours,respectively, enclose positive and negative flux regions with $|B_z|\ge 700$ G). Although these footpoints are located in downflows, they are relatively bright ($I/\langle I \rangle = 1.4-1.6$).}\label{fig:darklane}
\end{figure}

Similar observational features in emerging flux regions can also be found in our simulated EFR. Fig.~\ref{fig:darklane} shows snapshots of the vertical component of the magnetic field ($B_z$, sampled at $\tau_{\rm 500}=0.1$) and continuum intensity distribution near a darkening within the EFR. The darkening is predominantly associated with upflowing plasma ($v_z$ up to $1$ km s$^{-1}$) whereas the flanking bright grains at the ends of the darkening are associated with the footpoints of the rising flux bundle. In our EFR simulation, however, there also exist transient darkenings (of comparable spatial extent and lifetime) which are predominantly associated with downflows having speeds of about $0.5-1$ km s$^{-1}$. In addition, even those darkenings which are initially associated with upflows eventually (within $5-10$ minutes) become associated with downflows. This transition reflects that fact that although low entropy material can overshoot into the photosphere, it eventually overturns and forms new downflow lanes in the granulation pattern (mass conservation entails that most upflowing plasma must overturn within a few pressure scale heights).

\subsubsection{Transient kilogauss horizontal fields}
\begin{figure}
\centering
\includegraphics[width=0.46\textwidth]{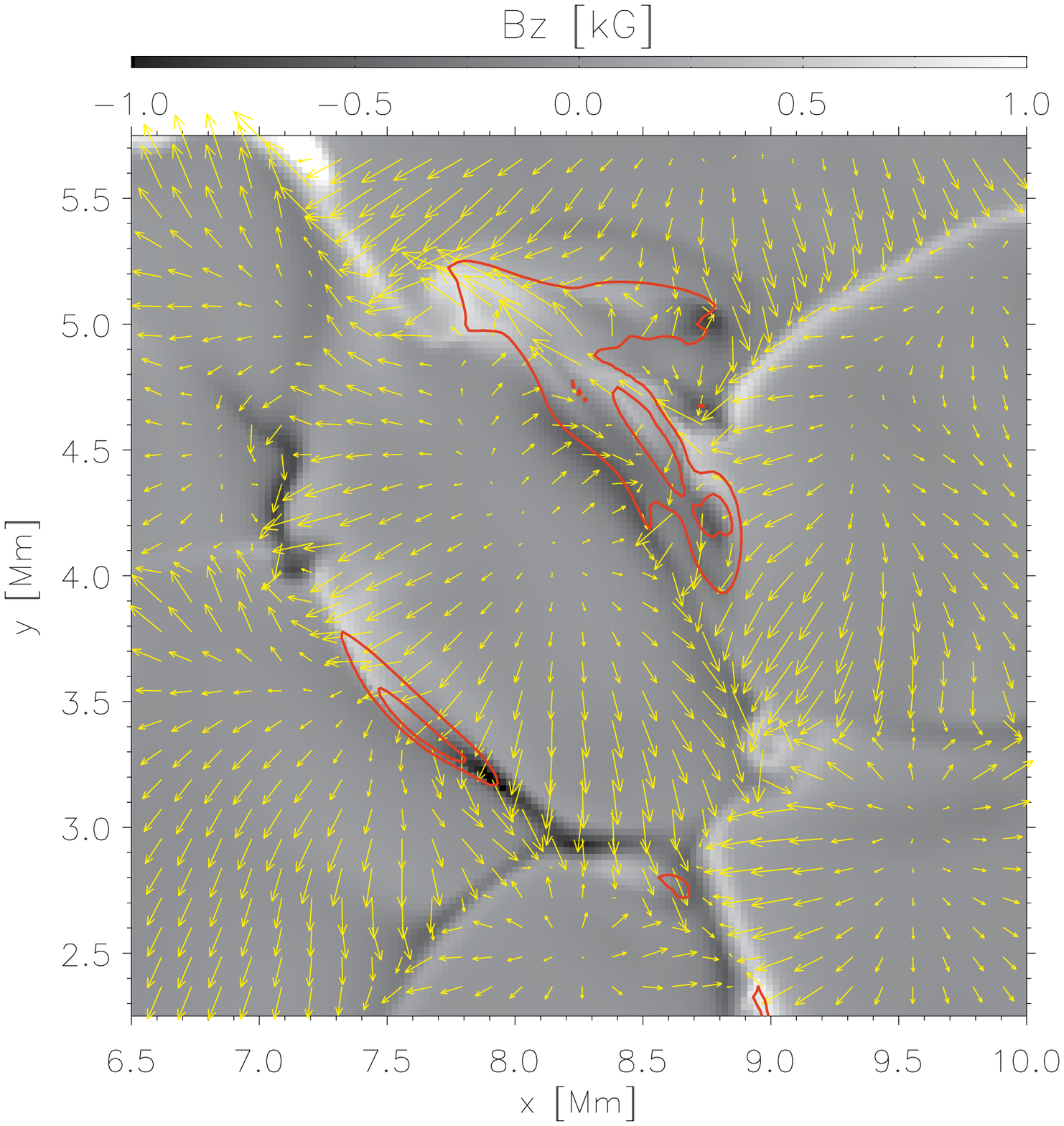}
\caption{Converging horizontal flows from neighboring granules can intensify surface horizontal fields to field strengths above $1$ kG. The vertical magnetic field ($B_z$, greyscale) is overplotted with the horizontal flow field (indicated by yellow arrows) and contours of $|B_{\rm hor}|=700$ G and $1200$ G. All quantities were sampled from the $\tau_{500}=0.1$ surface.}\label{fig:kilogauss_bhor}
\end{figure}

Measurements of horizontal photospheric fields outside of sunspot penumbrae have thus far resulted in sub-kG field strengths. For instance, ground-based observations of the quiet Sun with the Advanced Stokes Polarimeter (ASP) revealed internetwork horizontal fields with field strengths up to $600$ G~\citep{Lites:HIFs}. Similarly, ASP observations of EFRs by~\citet*{Lites:EmergingFieldsVector} and~\citet*{Kubo:EmergingFluxRegion} led the authors to conclude that flux emerges initially as horizontal structures with sub-kG field strengths, and that only after the emerged fields become vertical do super-kG fields exist.

Recent observations of ubiquitous horizontal fields pervading the photospheric surface~\citep{Harvey:SeethingHorizontalFields,Orozco:QuietSunFields,Lites:HinodeHorizontalFields} have revealed average horizontal fields on the order of tens of G. The lack of a dependence of the mean horizontal field strength with solar latitude is strongly suggestive of the existence of a surface dynamo, which has in fact been numerically modeled with the MURaM code~\citep*{Voegler:SolarSurfaceDynamo}. In any case, ubiquitous horizontal fields resulting from the surface dynamo are limited to field strengths much weaker than $1$ kG. In the following, we report that transient super-kG horizontal fields can exist within emerging flux regions. 

Figure~\ref{fig:kilogauss_bhor} shows the surface vertical field strength ($B_z$, grey-scale), horizontal velocity field (arrows) and horizontal field strength (red contours indicate $|B_{\rm hor}|$ = $700$ G and $1200$ G) within a portion of the simulated EFR (run B) at $t=60$ min. In this example, we find a couple of strong concentrations of predominantly horizontal field at the edge of neighboring granulation cells. The field strength in these concentrations can reach up to $1.5$ kG, which is comparable to the strong vertical fields inferred from model inversions of Stokes profiles observed in the quiet Sun~\citep{Lin:kGFields,SanchezAlmeida:MISMA2,DominguezCerdena:InternetworkMagneticFields2003}. Taking the mean density at the photospheric base to be $\langle \varrho \rangle = 2.6\times10^{-7}$ g cm$^{-3}$, and the surface flow speed of granular motion to be $v=4$ km s$^{-1}$, the photospheric equipartition field strength is $B_{\rm eq} = (4\pi\langle \varrho \rangle)^{1/2}v  = 700$ G. This means that a field strength of $1.2$ kG significantly higher than the equipartition value.

The mechanism for the intensification of horizontal fields to super-kG strengths is related to the convective intensification process that occurs for vertical flux concentrations in the sense that mass drainage from the loop is important. Consider a granulation-scale $\Omega$-loop at the surface, with the footpoints of the loop located in the intergranular lanes. As a result of mass drainage through the footpoints of the loop, the horizontal portions of the loop become evacuated ($\varrho/\bar{\varrho}\approx 0.5-0.8$) and allow two effects to work together to intensify the field. First, the decrease in the internal gas pressure allows the colliding outflows from adjacent granules to compression the fluid. Second, as the colliding flows encounter each other and are deflected to the side, the horizontal portions of the small-scale loop are stretched. These two effects are basically the first and second terms of Eq. (\ref{eqn:induction_lagrangian}) respectively. As is the case for vertical flux concentrations, the partial evacuation of the loop results in a local Wilson depression, so that the $\tau_{500}=0.1$ surface is closer to $z=0$ than $z=160$ km. This allows photospheric lines to probe into slightly deeper layers and provides an apparent intensification.

Although we have presented an example of how kG horizontal fields may be formed in an EFR, we do not claim that they are common. Rather, we are merely pointing our the possibility that, under some conditions, such structures may indeed exist and perhaps be detected.

\section{Discussion}
\label{sec:discussion}

The results in this paper have interesting implications for our understanding of the subsurface evolution of emerging flux as well 
as their observational diagnostics near the solar photosphere. By comparing our model results with the new observational data from 
the Hinode Solar Optical Telescope, we are able to discern the physical processes underlying many observed properties of emerging flux regions.

For instance, the mixed field morphology in the interior of EFRs is reproduced by the simulations, which reveal the serpentine nature
magnetic field lines as they attempt to rise towards the surface~\citep{StrousZwaan:SmallScaleStructure,Pariat:ResistiveEmergence}. From this work, we find that the undulation is naturally explained by the interaction of convective downdrafts and the rising magnetic field lines. Although mixed-polarity field exists within an EFR, the pattern has a certain order to it, namely that flux of one polarity tends to stream towards one direction, whereas the opposite polarity flux migrates in the opposite direction~\citep{StrousZwaan:HorizontalDynamics,Bernasconi:EmergingFlux}. At each of the two opposing edges of the the EFR, flux of one particular sign tends to coalesce. When sufficient flux comes together, the formation of solar pores and perhaps sunspots can occur. In a striking example of flux emergence observed by SOT (see Fig.~\ref{fig:march8ef}), solar pores
formed within one hour after the beginning of flux emergence. Due to the diminishing time-step imposed by the CFL criterion (mainly
from the high Alfv\'en speed within strong field regions), our simulations
have not yet progressed to the stage of pore formation. However, we do find a trend in both simulations runs of surface flux
coalescence, leading to some individual concentrations with up to $10^{19}$ Mx. It is our aim in future studies to advance these (or similar) simulations to
the stage that pores form as a natural consequence of flux emergence~\citep[as opposed to the simulations of~][which started with a coherent 
vertical flux tube as an initial condition]{Cameron:Pores}. 

One interesting observational feature is the presence of supersonic downflows ($v_z < -6$ km s$^{-1}$) at some flux `cancellation' sites, which can be found in both our models and the Hinode SP observations (see section~\ref{subsubsec:supersonic_downflows} and~\citeauthor{Shimizu:HighSpeedMassDownflows},~\citeyear{Shimizu:HighSpeedMassDownflows}). A possible cause of supersonic downflows at flux cancellation sites is revealed by the simulations to be a consequence of the relaxation of inverse-U loops (resulting from changes in connectivity). The magnetic tension in these loops
is sufficiently strong to accelerate plasma to supersonic ($M_s>1$), but sub-Alfv\'enic ($M_A < 1$), speeds. For the example shown in Fig.~\ref{fig:flux_cancellation}, the predominant cause of surface flux removal is actually the retraction of inverted U-loops (as opposed to magnetic diffusion).

Following the observational study by~\citet{Nagata:ConvectiveCollapse} in which they showed a clear example of convective collapse/intensification at work,
we present further evidence of how this process leads to the amplification of surface fields and to the formation of bright points~\citep[see also][]{Voegler:MURaM}. In section~\ref{subsubsec:convective_collapse},
we identified a clear case of this process happening to a magnetic flux concentration in our simulated EFR.

In the context of local helioseismology, the results from this study suggest that in the near-surface layers, EFR regions are likely to be
in the form of a magnetic flux sheet permeating the layers in the convection zone immediately beneath what magnetograms reveal as the surface
manifestations of EFRs. The spread of the tube into sheet-like structure is due to the strong horizontal expansion it experiences in the 
top few pressure-scale heights of the convection zone. This behavior, which was originally predicted by~\citet*{SpruitTitleVB:WeakMixedPolarityBackground}, can also be found in previous idealized simulations of flux tubes emerging from the convection zone to the solar atmosphere~\citep[e.g.][]{Magara:2.5D,Archontis:EmergenceIntoCorona}.

Another finding which may be relevant for the local helioseismology of EFRs is the presence of a~\emph{magnetic inversion layer} immediately 
above the photospheric base (i.e. above $\tau_{\rm 500}=1$) of upflow granules (see section~\ref{subsubsec:inversion_layer}). This layer extends throughout the horizontal extent of the simulated EFR and is present throughout the emergence process. It is associated with low entropy plasma which has recently emerged at the surface and experienced rapid radiative cooling. The result of the radiative cooling is an enhancement of the plasma density 
and magnetic field strength relative to the plasma immediately below. The thickness of the transition layer is comparable to the thermal boundary layer (a few
tens of km) and the relative enhancements of $\varrho$ and $|B|$ are of order $10^{-1}$. Thus, the magnetic inversion layers above granules may practically
 act as sharp boundaries for propagating magneto-acoustic waves, the effects of which may need to be carefully considered. 


\begin{acknowledgements}
We thank the SOT/FPP team for making high quality observations of magnetic flux emergence. This work was supported by NASA contract NNM07AA01C at LMSAL. Hinode is a Japanese mission developed and launched by ISAS/JAXA, collaborating with NAOJ as domestic partner, and NASA and STFC (UK) as international partners. Science operation of Hinode is conducted by the Hinode science team organized at ISAS/JAXA. Postlaunch operation support is provided by JAXA and NAOJ (Japan), STFC (UK), NASA, ESA, and NSC (Norway). The numerical simulations were carried out on the computing facilities at the Gessellschaft f\"ur wissenschaftliche Datenverarbeitung mbH G\"ottingen (GWDG). M. C. M. Cheung would like to thank fellow participants of the Flux Emergence Workshop 2007 at the University of St. Andrews for stimulating discussions. We also thank the anonymous referee for insightful comments which helped to clarify the presentation of results.
\end{acknowledgements}

\bibliographystyle{apj}

\end{document}